# Direct and Heterodyne Detection of Microwaves in a Metallic Single Wall Carbon Nanotube

F. Rodriguez-Morales[a], R. Zannoni[a], J. Nicholson, M. Fischetti and K.S. Yngvesson[b]

*Department of Electrical and Computer Engineering, University of Massachusetts, Amherst, MA 01003, USA*

J. Appenzeller

*IBM T. J. Watson Research Center, P.O. Box 218, Yorktown Heights, New York 10598, USA*

[a] These authors contributed equally to this paper; [b] email yngvesson@ecs.umass.edu

ABSTRACT. This letter reports measurements of microwave (up to 4.5 GHz) detection in metallic single-walled carbon nanotubes. The measured voltage responsivity was found to be 114 V/W at 77K. We also demonstrated heterodyne detection at 1 GHz. The detection mechanism can be explained based on standard microwave detector theory and the nonlinearity of the DC IV-curve. We discuss the possible causes of this nonlinearity. While the frequency response is limited by circuit parasitics in this measurement, we discuss evidence that indicates that the effect is much faster and that applications of carbon nanotubes as terahertz detectors are feasible.

The availability of Single Wall Carbon Nanotubes (SWNTs) has stimulated considerable recent exploration of different ideas for use of SWNTs in new electronic devices[1]. SWNTs have different electronic band structures depending on their chirality, and can be either metallic or semiconducting[2]. Much of the exploratory effort so far has been concentrated on the semiconducting version of SWNTs (s-SWNTs), in particular with the prospect of developing high performance Carbon Nanotube Field Effect Transistors (CN-FETs)[3,4]. Other applications that have been proposed are to detectors for microwave or terahertz frequencies. Schottky barriers exist at the contacts of semiconducting SWNTs[5,6], and were fabricated and analyzed for use as terahertz detectors by Manohara *et al.*[7] Experimental results were recently published by Rosenblatt *et al.*[8] demonstrating detection of microwaves up to 50 GHz, as well as by Pesetski et al.[9] who measured heterodyne detection with flat frequency-dependence up to 23 GHz. These references[7,8,9] all used the s-SWNT-FET configuration. Metallic SWNTs (m-SWNTs) also have considerable potential for detector applications, and one of us (KSY) recently proposed a very fast terahertz detector based on the hot electron bolometric (HEB) effect[10]. In the present letter we report experimental results for a device using an m-SWNT that detects microwaves in the low GHz range, based on a traditional IV-curve nonlinearity The device described here operates both as a direct (DC output) detector and as a heterodyne detector (difference frequency output up to at least 200 MHz). In this paper we will discuss the experimental results and interpret these in terms of the detection mechanisms involved. We also discuss the potential of this type of detector for application at terahertz frequencies.

SWNTs used in our study were grown using laser ablation [11]. CNTs with diameters between 0.6 nm and 1.5 nm are spun from solution onto a p+-doped silicon substrate covered with 100 nm of silicon oxide. Contact strips of width 350nm were made with 20 nm of Ti followed by 100 nm of Au, and were connected to 80μm x 80μm contact pads. The length of the tubes between contacts is known to be in the range of 300nm to 500nm. The silicon chip was placed in a small copper enclosure (with a metallic cover) to isolate it from external radiation, see Figure 1. The contact pads were connected by wire bonds to (1) a microstrip transmission line that was in turn connected to a standard coaxial connector installed in the side of the enclosure; and (2) the ground plane of the enclosure. The silicon substrate was left electrically insulated in order to minimize parasitic reactances. The assembly was placed in a liquid helium vacuum dewar and pumped to a good vacuum for at least one day in order to remove most of the surface contaminations on the CNT. A well shielded stainless steel coaxial cable makes the sample accessible from the outside of the dewar. We used a programmable DC power supply (Keithley) to provide a voltage source bias to the device through the coaxial cable. The DC supply also measured the DC voltage and current, and these were read by a computer for further processing. Microwave sources (Agilent) were also fed to the coaxial cable, and different sources (DC and microwave) were separated through the use of commercial bias tees.

It is well-known that Ti/Au contacts yield a contact resistance that is usually quite high and strongly depending on the nanotube diameter[12,13,14]. The devices used in our study had contact resistances that were in the range of a few hundred kΩ to a few MΩ. It is also known that the conductance of such CNTs shows a "zero-bias anomaly"[15], i.e. the differential conductance (dI/dV) plotted as a function of bias voltage (V) shows a dip at low values of V with a width of about +/- 400 mV[16]. This presents a nonlinearity in the IV-curve (Figure 2) that we exploited for microwave detection.

The zero-bias anomaly "dip" is also evident from the additional plot of dI/dV in Figure 2. This dip deepens as the temperature is decreased (the curves shown in Figure 2 were taken at 77K). At larger voltages the IV-curve shows a linear dependence between current and bias with a slight decrease in dI/dV for the highest voltage range. Except for the zero-bias anomaly, the IV-curve can thus be assumed to be due to a (roughly) constant contact resistance, that is almost independent of the temperature. Evidence from other metallic CNTs[17] indicates that the electrons have mean free paths of about 1µm; thus in our tubes they travel ballistically from contact to contact. The zero-bias anomaly is usually ascribed to the very strong electron-electron Coulomb interactions in one-dimensional conductors that necessitates treating the electrons as a collective, plasmon-like, medium known as a "Luttinger liquid" ("LL")[18]. Tunneling from the contacts into the LL is suppressed at low temperatures, which explains why the conductance approaches zero. It has been suggested that the behavior of the conductance in the entire temperature range from 4 K to 300 K can be better explained as being due to a combination of effects, the LL effect, and that of interfacial barriers at the contacts[14]. The LL effect is expected to be important only in the lowest temperature range. As made clear in the paper mentioned above[14], a complete understanding of the contacts between the one-dimensional m-SWNTs and a 3-D metal is not yet available.

As microwaves were applied to the SWNT at 77K, we recorded a change in the device DC current (ΔI), and plotted this versus DC bias voltage (Figure 3 (a)). This recording was done by measuring the voltage across a series resistance with a lock-in amplifier, while square wave modulating the microwave source. The DC power supply was still configured as a voltage source. The microwave reflection coefficient (S11) was also measured with an automatic network analyzer, see Figure 4. This particular recording was obtained for a CNT with resistance of a few MΩ, but similar results were obtained for in total three samples. A resonance is seen at about 1.28 GHz, which we interpret as being due to the combined effect of the bond wires, the contact pads and the connecting strips, situated on top of the oxide and the doped silicon chip. The equivalent circuit shown in the inset of Figure 4 was used to produce a good fit to the magnitude of S11, as shown in Figure 4. For this fit we used the full S11 data, including the real and imaginary parts (not shown explicitly). We also used the model to predict the measured detected change in current versus frequency, plotted in Figure 4 for two different microwave power levels. The detected signal is essentially independent of frequency below the resonance, indicating that the effect of any parasitic reactance is negligible at frequencies below about 900 MHz. Above the resonance frequency, the response falls off by 12 dB per octave, in good agreement with the model. The highest frequency at which we detected the signal was 4.5 GHz, limited by the sensitivity of our measurement system. Given that s-SWNTs detected microwaves up to 23GHz and 50GHz, respectively[6,7], it is reasonable to assume that the detection effect we report here for m-SWNTs will extend to similarly high frequencies, once parasitic effects have been minimized.

At 77 K, the detected DC current change (ΔI) depends linearly on the microwave power (a "square law detector") up to a power of about 0.02 mW; the detected current change then decreases smoothly after passing a maximum at about 1 mW, see Figure 3(b). The linear current responsivity at low MW powers was found to be $S_I = \Delta I/P_{MW} = 455$ µA/W, based on the measured output power at the microwave source ($P_{MW}$). The response was also detectable at 300 K, and with much higher (40x) responsivity at 4K. The behavior of this detector at the lowest temperatures needs to pursued through further measurements and will be presented in a future paper. The current responsivity can be converted to a voltage responsivity ($S_V$) by multiplying with the device resistance, 250 kΩ, yielding $S_V = 114$ V/W. Higher resistance CNTs have lower values for $S_I$, roughly in inverse proportion to the resistance, and therefore have about the same $S_V$. Using standard small-signal microwave detector theory[19] we can calculate the current responsivity from the following expression:

$$\Delta I = (1/4)*(d^2I/dV^2)*V_{MW}^2 \qquad (1)$$

Here, $V_{MW}$ is the peak microwave voltage. The factor $d^2I/dV^2$ was calculated from the measured IV-curve, and is compared with $\Delta I$ in Figure 3 (a). The small oscillations in the plot of $d^2I/dV^2$ are an artifact of the measurement method caused by the finite steps produced by the voltage source. The linear dependence of $\Delta I$ on MW power in the small-signal regime, as shown in the inset of Figure 3(b), indicates that Eq. (1) applies. Further, the bias voltage dependence of $\Delta I$ agrees well with that of $d^2I/dV^2$ (Figure 3(a)). For a microwave power of 10 µW we use Eq. (1) to estimate $\Delta I$ in the range 5nA to 20nA, depending on the detailed assumptions made about the values of the equivalent circuit elements in Figure 4. The measured value is 5nA, and this quantitative agreement within expected error bars gives further strong support to the interpretation that the detector operates as a standard microwave detector with a response that can be predicted from its IV-curve. For higher microwave powers, the small signal approximation becomes invalid, and the response becomes nonlinear, as is clear from Figure 3(b). We note that since the transport in the m-SWNT is ballistic, the entire nonlinearity of the detector is due to the contact resistance.

We next demonstrated *heterodyne* detection in the same SWNT by connecting it to two microwave sources with different microwave frequencies $f_1$ (designated as the "Local oscillator, LO") and $f_2$ ("RF or signal frequency"), while measuring the output power (or voltage) at the difference frequency (IF), ($|f_1-f_2|$). The IF power seen on a spectrum analyzer (inset in Figure 5) was essentially independent of the IF frequency up to 200 MHz. Detecting a higher IF was not possible due to the properties of the bias tees used. The detected IF voltage response versus DC bias voltage is shown in Figure 5.

For the data plotted in Figure 5 we used a more sensitive method of detecting the IF on a lock-in amplifier. The reference voltage for the lock-in amplifier was created by employing a separate commercial microwave mixer to mix $f_1$ and $f_2$, see e.g. Sazonova et al.[20]. Typical frequency combinations used were $f_1$ and $f_2$ near 1 GHz, with an IF of 50 kHz. Again, parasitic circuit elements on the chip decreased the mixer efficiency for $f_1$ and $f_2$ above 1 GHz. As for the direct detection case, the response follows $d^2I/dV^2$ when the bias voltage is varied (compare Figure 3(a)). This indicates that the heterodyne detection mechanism is attributable to the IV-curve using standard mixer theory.

We estimate a total mixer conversion loss to a 50Ω IF amplifier of 95dB, much of which is due to the high mismatch loss (60 dB total) to a device with 250 kΩ resistance.

Lower resistance SWNTs[3,4] would show lower mismatch loss as mixers. Note, however, that the higher resistance SWNTs show very good performance as *direct* detectors.

We now want to further discuss some of the implications of our experimental data. We have shown that sensitive detection of microwaves is possible in an m-SWNT. The detector response follows standard microwave detector theory, based on the zero-bias anomaly nonlinearity (ZBA) in the IV-curve. The origin of the ZBA has been much discussed, and this discussion is still ongoing. Especially interesting is to understand how the character of the electron transport changes as the temperature and the bias voltage are changed. The Luttinger liquid (LL) theory has been invoked to explain the ZBA, with the main experimental evidence for this theory being provided by the power-law dependence of the conductance on $eV/kT$[18]. Further microwave detector studies would be useful for exploring this problem. The fact that the microwave detection response is well predicted by the DC IV-curve indicates that whatever effect that causes the ZBA, it operates at speeds up to at least 4.5 GHz. This frequency limit is presently set only by the parasitics of the circuit, not the SWNT. It would be of great interest to extend the studies of coupling high frequency fields to SWNTs from the gigahertz range to the terahertz range in order to explore the intrinsic speed of the SWNT. Resonances in the LL are predicted to occur at frequencies in the terahertz range for the length of SWNTs studied here[22]. It appears promising to extend the present study and explore potential terahertz detectors based on m-SWNTs[10].

The authors thank Eric Polizzi and Eyal Gerecht for stimulating discussions. This work was supported by NSF grant ECS-0508436 for Nanoscale Exploratory Research.

# FIGURE CAPTIONS.

Figure 1. The experimental fixture used in this work.

Figure 2. Measured IV-curve for a SWNT at 77 K (right scale); dI/dV based on the IV-curve (left scale).

Figure 3. (a) Detected DC current change ($\Delta I$; points connected with line segments) due to microwave signal at 900 MHz and 77K, compared with $d^2I/dV^2$ (fulldrawn), input power -20 dBm; (b) The DC current change ($\Delta I$) at negative peak of Figure 3(a), as a function of microwave power. Inset: expanded view of 3(b).

Figure 4. Microwave frequency dependence of the detected DC current change (at two power levels; left scale) and the magnitude of the reflection coefficient S11 ( right scale; dB units), compared with the data predicted from the circuit model. Inset: Circuit model.

Figure 5. The detected IF voltage in the heterodyne detection mode, plotted versus DC bias voltage at 77K; $P_{LO}$ = -10 dBm; $P_{RF}$= -20 dBm. LO and RF frequencies close to 900MHz. Inset shows the IF response on a spectrum analyzer.

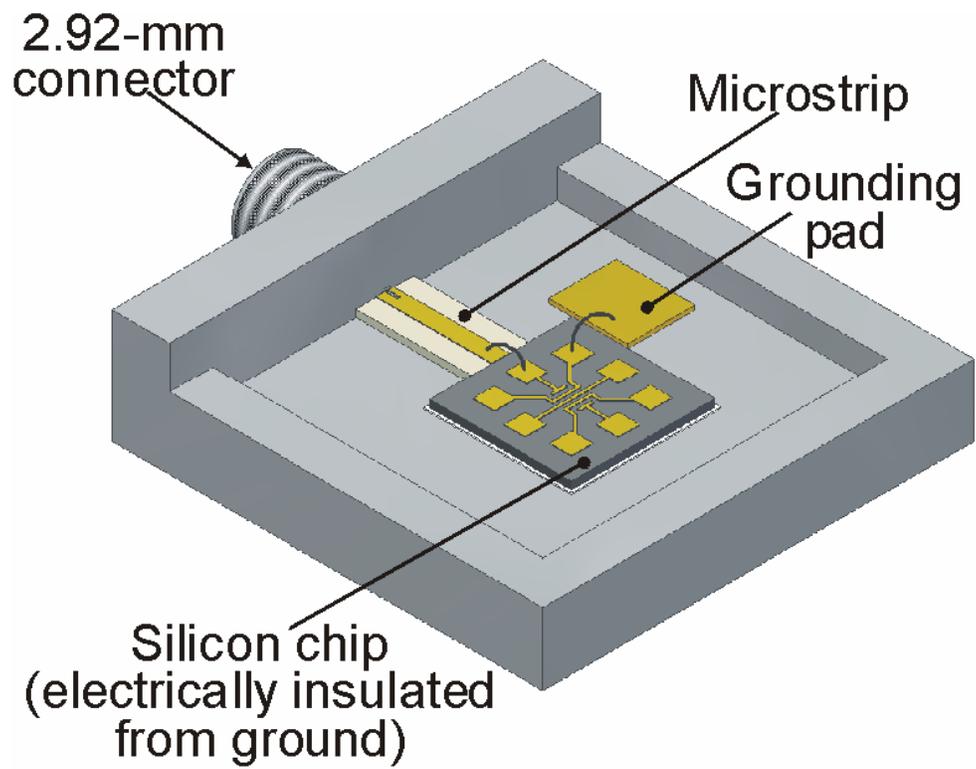

FIGURE 1

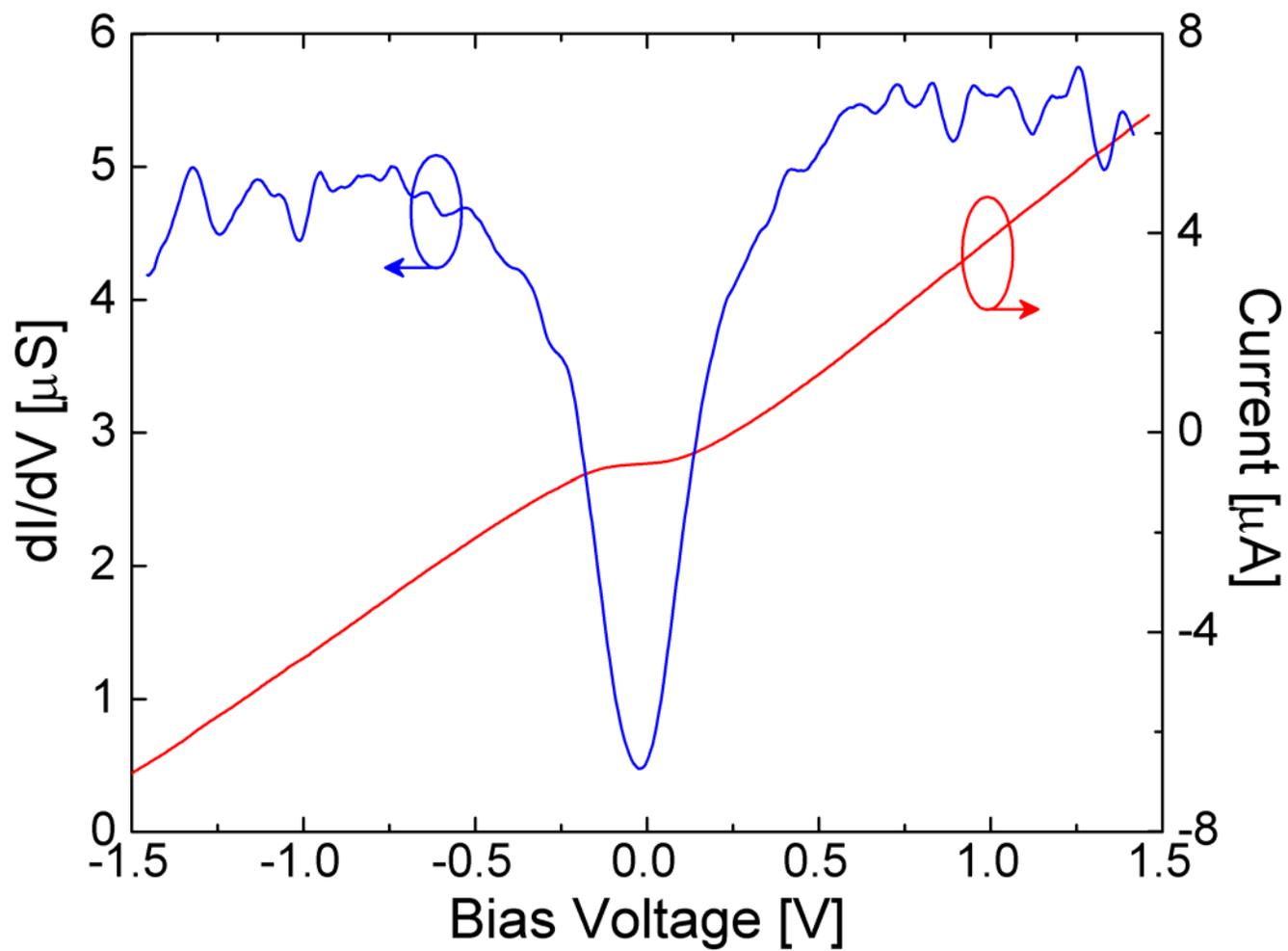

FIGURE 2

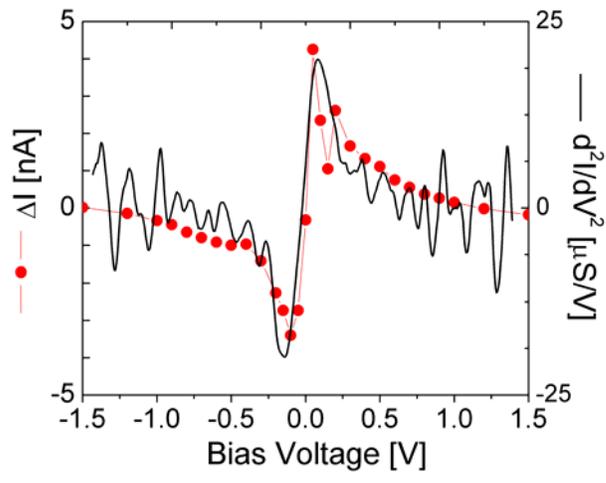 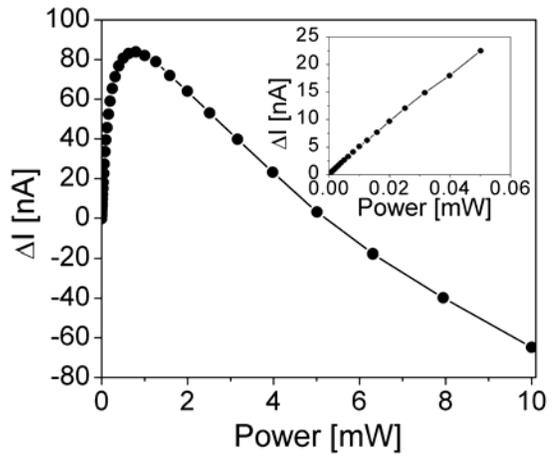

FIGURE 3

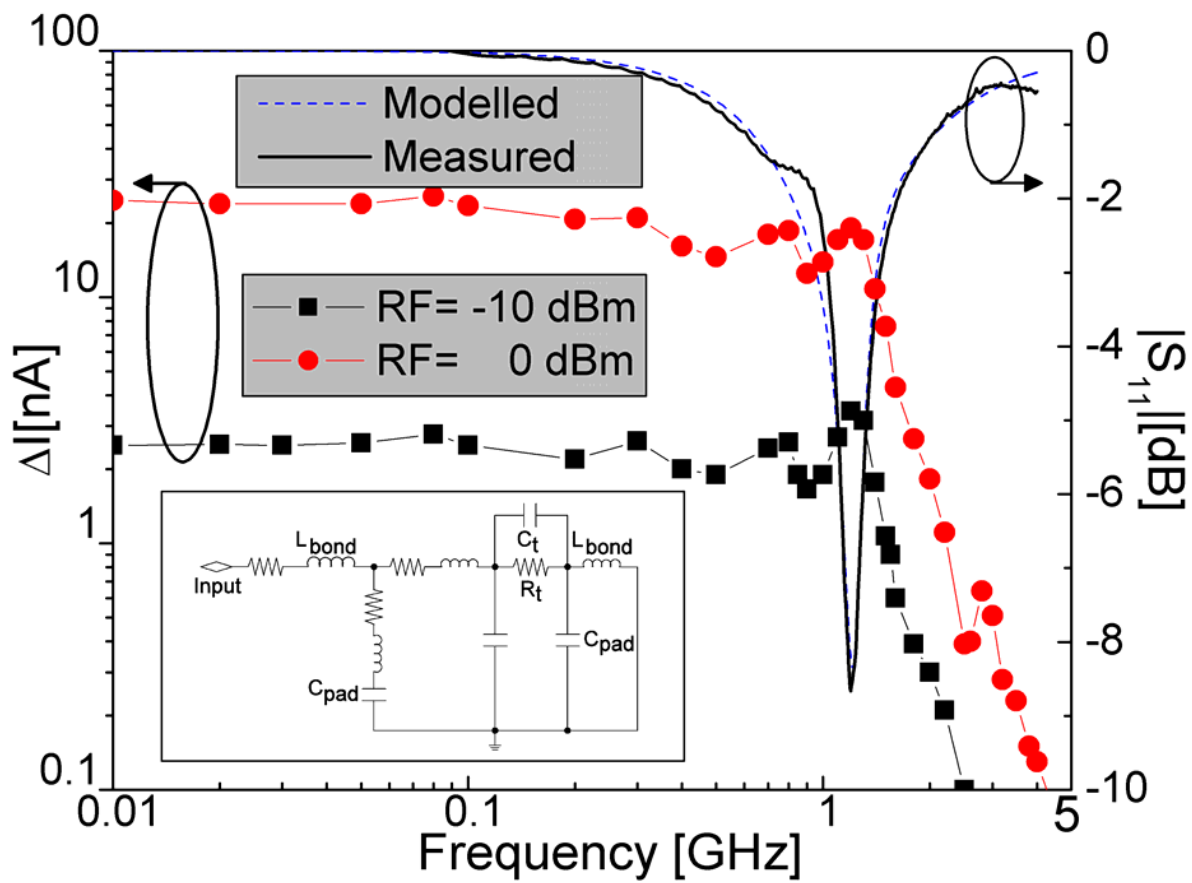

FIGURE 4

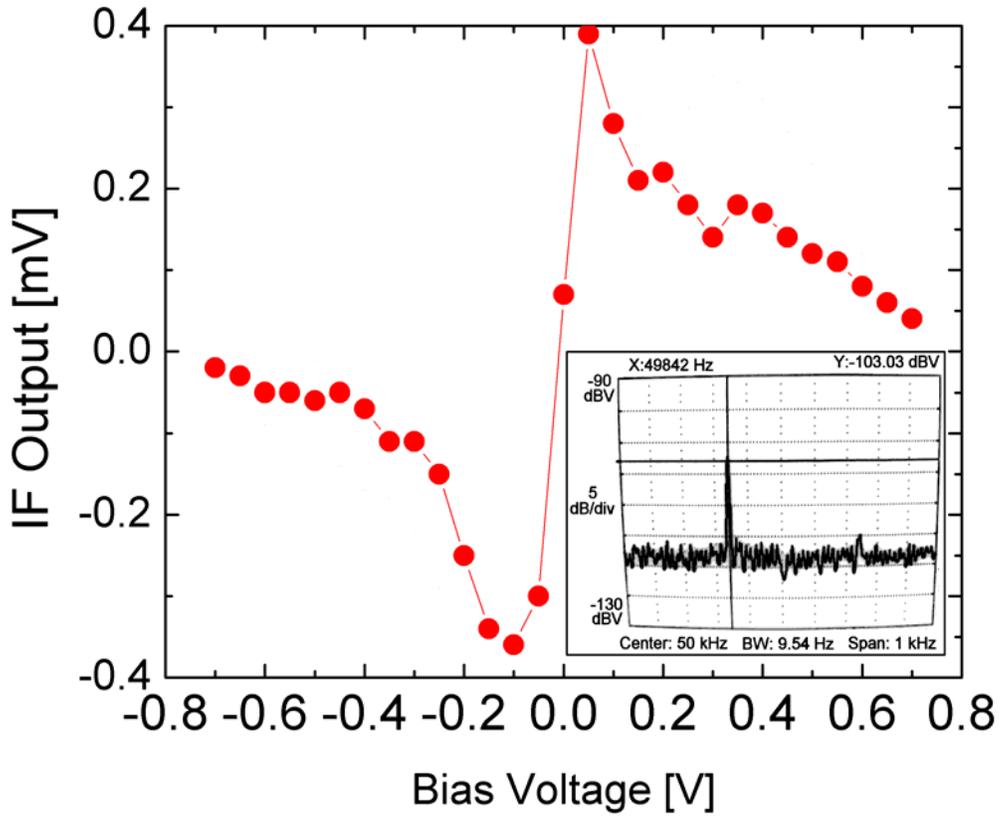

FIGURE 5